\pdfoutput=1
\documentclass[journal=jpclcd,manuscript=letter]{achemso}

\usepackage[version=3]{mhchem} 

\usepackage{amsmath,amssymb}
\usepackage{mciteplus}
\usepackage{graphicx}
\usepackage{bm}
\usepackage{float}
\usepackage{subfigure}
\usepackage{color}
\usepackage{mciteplus}
\usepackage{tabularx}
\usepackage{rotating}
\usepackage{lmodern}
\usepackage{physics}

\usepackage[dvipsnames]{xcolor}

\author{Siwei Wang}
\affiliation{Department of Chemistry and Biochemistry, University of Notre Dame, Notre Dame, Indiana 46556, United States}

\author{Liang-Yan Hsu}
\affiliation{Institute of Atomic and Molecular Sciences, Academia Sinica, Taipei 10617, Taiwan}
\alsoaffiliation{Department of Chemistry, National Taiwan University, Taipei 10617, Taiwan}
\alsoaffiliation{Physics Division, National Center for Theoretical Sciences, Taipei 10617, Taiwan}
\email{lyhsu@gate.sinica.edu.tw}

\author{Hsing-Ta Chen}
\affiliation{Department of Chemistry and Biochemistry, University of Notre Dame, Notre Dame, Indiana 46556, United States}
\email{hchen25@nd.edu}

\title[An \textsf{achemso} demo]
  {Robust Surface-Induced Enhancement of Exciton Transport in Magic-Angle-Oriented Molecular Aggregates}

\begin{document}


\begin{tocentry}
\includegraphics[width=2.0in,angle=0]{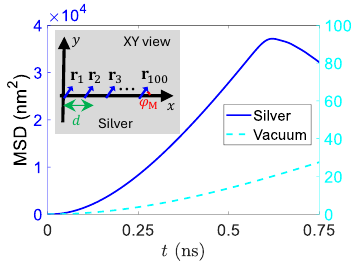}
\end{tocentry}

\begin{abstract}
    Exciton transport in molecular aggregates with magic-angle orientation is expected to be strongly suppressed due to their negligible dipole-dipole interactions. However, recent reports show that light-matter interactions can significantly enhance exciton transport attributed to the effective long-range coupling mediated by the photonic fields. To elucidate their interplay, we employ the macroscopic quantum electrodynamics framework to simulate exciton transport within a chromophore array arranged in a magic-angle configuration in proximity to a silver surface. Our results show a significant enhancement of the exciton diffusion coefficient that is robust across variations in chromophore-surface separation, intermolecular distance, and molecular transition frequency. Furthermore, based on the image-dipole method, we derive analytical expressions that agree well with numerical simulations, revealing the enhancement's origin in the near-field coupling term as induced by the radiative scattering at the metallic surface. More importantly, we observe non-trivial differences in the diffusion coefficient's scaling near metallic surfaces compared to free space. Our findings highlight the potential to control exciton transport by designing coupled exciton-photon systems and engineering the dielectric environments. 
\end{abstract}

\maketitle

Harnessing exciton transport in molecular systems has been a key goal of extensive theoretical and experimental investigation over the past decades. Beginning with Frenkel’s foundational model of localized excitations\cite{frenkel_1931}, now known as Frenkel excitons, many further theoretical developments have been inspired\cite{kenkre2006exciton}. The molecular aggregate model by Spano \emph{et al.} revealed the connection between aggregate morphology and optical properties\cite{spano_2014}. More advanced frameworks by Cao and Brumer {\it et al.} addressed environment-assisted transport and coherence effects\cite{cao_1994,pachn_2011,schinabeck_2016}. On the experimental front, ultrafast spectroscopic techniques have been developed by Fleming and Scholes {\it et al.} to probe quantum coherence and its effect on exciton transport, particularly in natural light-harvesting systems\cite{scholes_2011,schultz_2024}. Another related area of significant progress stems from studies on organic solar cells, which have significantly advanced the understanding and optimization of exciton diffusion for high-performance photovoltaic devices, with substantial contributions from researchers such as Meredith and Leo.\cite{riede_2021,riley_2022} These studies have profoundly deepened our understanding of excitation energy transport in molecular aggregates and led to new design principles of light-harvesting systems and organic solar cells\cite{scholes_2006,brdas_2016}.

Exciton transport efficiency in molecular aggregates is strongly dictated by the relative orientations of two chromophores, often characterized by the orientation factor in the rate of F\"{o}rster resonance energy transfer\cite{scholes_2003, somayaji_2025}. Notably, it has been shown that exciton transport can be strongly suppressed in molecular aggregates stacking with a ``magic-angle'' orientation and a large intermolecular distance ($d > 1~\mathrm{nm}$) due to simultaneous quenching of the primary exciton transport mechanisms.\cite{hestand_2018} On the one hand, the F\"{o}rster resonance energy transfer based on the long-range intermolecular dipole-dipole coupling vanishes at the magic angle ($\varphi_\mathrm{M}\approx54.74^\circ$). On the other hand, the Dexter electron exchange governing the short-distance transport becomes negligible at large separations\cite{scholes_2003}. However, intermolecular interactions are determined not only by the intrinsic properties of the molecules and their spatial arrangement but also by the surrounding dielectric environment. Particularly, near a metallic surface, the optical and transport properties of molecular aggregates can be significantly modified, leading to phenomena such as the Purcell effect\cite{rybin_2016} and Casimir–Polder (CP) force\cite{casimir_1948,klimchitskaya_2009}. Furthermore, under strong coupling conditions, the formation of hybrid light-matter quasiparticles known as polaritons can profoundly alter electron transfer and exciton transport processes\cite{rozenman_long-range_2018,xiang_intermolecular_2020,bhatt_enhanced_2021,berghuis_controlling_2022,yuenzhou_2022,balasubrahmaniyam_enhanced_2023,xu_ultrafast_2023,sandik_cavity-enhanced_2024,zhou_nature_2024,brawley_2025,ying_microscopic_2025}. In this context, therefore, it is intriguing to consider the interplay between these complex dielectric effects and magic-angle-oriented aggregates.
That being said, the conventional framework based on cavity quantum electrodynamics (CQED), while effectively capturing strong light-matter couplings\cite{agarwalla_2016,agarwalla_2019,mandal_2023,koessler_2025,taylor_2025}, cannot capture the effects of complex dielectric environments\cite{hsu_2025}.

To properly describe the effects of complex dielectric environments on exciton transport, one promising approach is to employ macroscopic quantum electrodynamics (MQED), which provides a rigorous quantization framework for electromag netic fields in arbitrary inhomogeneous, dispersive, and absorbing dielectric media~\cite{Gruner1996,Dung1998,Dung2000, Wang2019,wei_2024,tsai_2024}. This formalism offers a powerful theoretical framework for studying medium-assisted intermolecular interactions~\cite{Dung2_2002,Martin-Cano2011,Ren2017,Hsu2017,Ding2018,Varguet2021}. 
More recently, the MQED framework has been generalized to develop microscopic theories of the exciton-polariton model for multichromophoric systems~\cite{Wang2022_MC,Chuang2022_MC,chuang_2024,chuang_2024_2,hsu_2025}, which can accurately capture enhanced superradiance rates~\cite{chuang_2024_PRL} and emission power spectra~\cite{Wang_2024_NonHerm} of multichromophoric systems near metallic surfaces. 
Building on these promising results, we employ the MQED framework in this study to simulate the exciton transport dynamics of magic-angle-oriented molecular aggregates. Specifically, we investigate three key questions: (1) whether the presence of a metallic surface can enhance the exciton diffusion (even if the molecular transition frequency is not resonant with the plasmon polariton frequency); (2) how robust such enhancement can be with respect to variations in molecular transition frequency $\omega_\mathrm{M}$ and geometric configurations, such as intermolecular separation $d$, and the distance between the aggregate and the metal surface $h$; and (3) how we can quantify their impact on the transport properties, such as diffusion coefficient. 

We consider a chain of $N_\mathrm{M}$ chromophores coupled to (dressed) photons, either in free space or above a silver surface. Here, each chromophore is modeled as a two-level system comprised of the energetic ground state $\ket{\mathrm{g}_\alpha}$ and excited state $\ket{\mathrm{e}_\alpha}$ for $\alpha=1,\cdots,N_\mathrm{M}$.
For demonstration purposes, we set the molecular aggregate comprising $N_\mathrm{M} = 100$ methylene-blue chromophores with a transition frequency $\omega_\mathrm{M} = 1.864$ eV and a transition dipole moment $\mu_\mathrm{M}=3.8$ Debye~\cite{Chikkaraddy2016}. Due to their rigidity, these chromophores can be reasonably approximated as two-level systems, and their molecular vibrational degrees of freedom can be ignored for exciton transport simulations.

\begin{figure}[htbp] \includegraphics[width=0.85\textwidth]{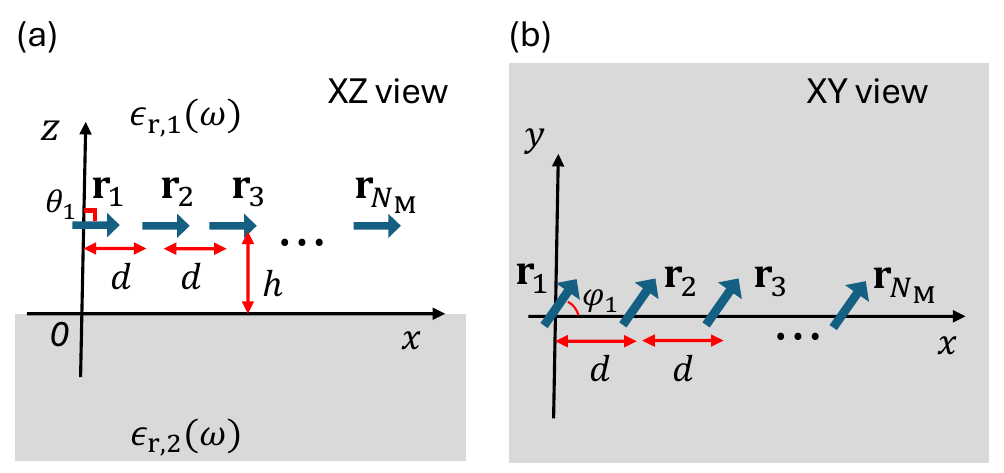}
\caption{Schematic diagrams of a magic-angle-oriented molecular aggregate above a silver surface. (a)
XZ view: The blue arrows represent the direction of the molecular transition dipole moments at positions ($\mathbf{r}_\alpha$ for $\alpha=1,\cdots,N$) in the side view. For all dipole moments, the polar angle relative to the z-axis is chosen to be $\theta_\alpha = 90^\circ $. The intermolecular distance between adjacent chromophores and the height of the chromophore above the silver surface are denoted as $d$ and $h$, respectively. The permittivity of vacuum and silver are represented by $\epsilon_\mathrm{r,1}(\omega)$ and $\epsilon_\mathrm{r,2}(\omega)$, respectively. 
(b) XY view: The blue arrows indicate the directions of the same molecular transition dipole moments in the top-down view. For all dipole moments, the azimuthal angles are chosen to be $\varphi_\alpha = \arccos(1/\sqrt{3}) \approx 54.74^\circ$. }
\label{Fig_Schematic}
\end{figure}

As illustrated in Figure~\ref{Fig_Schematic}, we assume that all chromophores are equally spaced by $d$ along the $x$-axis and placed at a uniform height $h$ above the silver surface (the $z=0$ plane). Explicitly, the position vector of the $\alpha$-th molecule is given by $\mathbf{r}_\alpha= \left( [\alpha-1] d,0,h \right)$.
We assume that the transition dipole moments of all chromophores are identical in amplitude, aligned parallel to the silver surface, and oriented at an azimuthal angle relative to the $x$-axis (explicitly $\boldsymbol{\mu}_\alpha= \mu_\mathrm{M}(\cos\varphi_\alpha,\sin\varphi_\alpha,0)$). Through the Letter, we choose the azimuthal angle to be $\varphi_\alpha=\varphi_\mathrm{M}=54.74^\circ$ for all $\alpha=1,\cdots,N_\mathrm{M}$, satisfying the magic-angle condition $1-3\cos^2\varphi_\mathrm{M}=0$.

Building on our previous studies based on the MQED framework\cite{chuang_2024,chuang_2024_2}, when the light-matter coupling is relatively weak, the reduced density matrix of molecular exciton states $\hat{\rho}_{\mathrm{M}}(t)$ follows the general form of the Gorini–Kossakowski–Sudarshan–Lindblad equation\cite{BreuerHeinz-Peter2007TToO} (see section S1 in the Supporting Information (SI)): 
\begin{align}
\frac{\partial}{\partial t} \hat{\rho}_{\mathrm{M}}(t) = -\frac{i}{\hbar} \left[ \hat{H}_{\mathrm{M}} + \hat{\mathcal{H}}_\mathrm{CP}^{\mathrm{Sc}} + \hat{\mathcal{H}}_\mathrm{RDDI}, \hat{\rho}_\mathrm{M}(t) \right] 
+ \sum_{\alpha,\beta}^{N_\mathrm{M}}  {\Gamma}_{\alpha\beta} \left( \hat{\sigma}_{\beta}^{(-)} \hat{\rho}_\mathrm{M}(t) \hat{\sigma}_{\alpha}^{(+)} - \frac{1}{2} \left\{ \hat{\sigma}_{\alpha}^{(+)} \hat{\sigma}_{\beta}^{(-)}, \hat{\rho}_\mathrm{M}(t) \right\} \right),
\label{Eq:GKSL_Eq}
\end{align}
where $\hat{\sigma}^{+(-)}_\alpha$ denotes the raising (lowering) operators for the $\alpha$-th chromophore (i.e., $\hat{\sigma}^{+(-)}_\alpha \ket{\mathrm{g_\alpha(e_\alpha)}}= \ket{\mathrm{e_\alpha(g_\alpha)}}$) and $[\hat{O}_1, \hat{O}_2]= \hat{O}_1  \hat{O}_2 - \hat{O}_2\hat{O}_1 $, $\{ \hat{O}_1, \hat{O}_2 \} = \hat{O}_1  \hat{O}_2 + \hat{O}_2\hat{O}_1 $. Here $\hat{H}_\mathrm{M}$, $\hat{\mathcal{H}}_\mathrm{CP}^{\mathrm{Sc}}$, and $\hat{\mathcal{H}}_\mathrm{RDDI}$ represent the bare Hamiltonian of $N_\mathrm{M}$ chromophores, the CP potential, and the resonant dipole-dipole interactions (RDDI), respectively:
\begin{align}
\label{Eq:Molecular_Hamiltonian}
& \hat{H}_\mathrm{M} = \sum_{\alpha=1}^{N_\mathrm{M}} \hbar \omega_\mathrm{M} \hat{\sigma}^{(+)}_\alpha \hat{\sigma}^{(-)}_\alpha , \\
&\hat{\mathcal{H}}_\mathrm{CP}^{\mathrm{Sc}} = \sum_{\alpha=1}^{N_\mathrm{M}} \Delta_{\alpha}^{\text{Sc}} \, \hat{\sigma}^{(+)}_\alpha \hat{\sigma}^{(-)}_\alpha , \\
& \hat{\mathcal{H}}_\mathrm{RDDI} =  \sum_{\alpha,\beta (\alpha\neq \beta) }^{N_\mathrm{M}} V_{\alpha\beta} \, \hat{\sigma}^{(+)}_\alpha \hat{\sigma}^{(-)}_\beta. 
\end{align}
In the MQED framework, the CP potential is associated with diagonal energy shifts in the presence of a dielectric environment: 
\begin{equation}\label{Eq:Interaction_Definition}
     \Delta_{\alpha}^{\text{Sc}} = \mathcal{P} \int_{0}^{\infty} d\omega \frac{\omega^2}{\pi \varepsilon_0 c^2}  {\boldsymbol{\mu}_{\alpha} \cdot \mathrm{Im}\overline{\overline{\mathbf{G}}}_{\text{Sc}}(\mathbf{r}_\alpha,\mathbf{r}_\alpha,\omega) \cdot \boldsymbol{\mu}_{\alpha}} \left( \frac{1}{ \omega + \omega_\mathrm{M} } -\frac{1}{\omega - \omega_\mathrm{M} } \right),
\end{equation}
where $\mathcal{P}$ denotes the Cauchy principal value, the RDDI characterizes the off-diagonal coupling strength:
\begin{equation}\label{Eq:RDDI_Definition}
    V_{\alpha\beta} = \frac{- \omega_\mathrm{M}^2}{ \epsilon_0 c^2}  \boldsymbol{\mu}_{\alpha} \cdot \mathrm{Re} \overline{\overline{\mathbf{G}}}(\mathbf{r}_\alpha,\mathbf{r}_\beta,\omega_\mathrm{M}) \cdot \boldsymbol{\mu}_{\beta},
\end{equation}
and the generalized dissipation rate is given by: 
\begin{equation}\label{Eq:Gamma_Definition}
    \Gamma_{\alpha\beta} = \frac{2\omega_{\mathrm{M}}^2}{\hbar\varepsilon_0 c^2} \boldsymbol{\mu}_{\alpha} \cdot \mathrm{Im}\overline{\overline{\mathbf{G}}}(\mathbf{r}_{\alpha},\mathbf{r}_{\beta},\omega_{\mathrm{M}}) \cdot \boldsymbol{\mu}_{\beta}.    
\end{equation}
In the above expressions, the propagation of the (dressed) photons from $\mathbf{r}_\beta$ to $\mathbf{r}_\alpha$ is captured by the dyadic Green's function $\overline{\overline{\mathbf{G}}}(\mathbf{r}_\alpha,\mathbf{r}_\beta,\omega_\mathrm{M})$ which satisfy the macroscopic Maxwell's equation,
\begin{align}
    \left( \frac{\omega_\mathrm{M}^2}{c^2}\epsilon_\mathrm{r}(\mathbf{r}_\alpha,\omega_\mathrm{M}) - \nabla \times \nabla \times \right)\overline{\overline{\mathbf{G}}}(\mathbf{r}_\alpha,\mathbf{r}_\beta,\omega_\mathrm{M}) = - \delta(\mathbf{r}_\alpha-\mathbf{r}_\beta) \mathbf{\overline{\overline{I}}}_3,
\label{Eq:Green_Fun_full}
\end{align}
where $\delta(\mathbf{r}_\alpha-\mathbf{r}_\beta)$ is the three-dimensional Dirac delta function, and $\mathbf{\overline{\overline{I}}}_3$ is the $3\times 3$ identity matrix. 
In free space (i.e. $\epsilon_\mathrm{r}(\mathbf{r}, \omega)=1$), the dyadic Green's function in eq~\ref{Eq:Green_Fun_full} can be solved analytically by
\begin{align}
    \overline{\overline{\mathbf{G}}}_0(\mathbf{r}_\alpha,\mathbf{r}_\beta,\omega_\mathrm{M})=&
    \frac{e^{ik_0 R_{\alpha\beta}}}{4\pi R_{\alpha\beta}}
    \left\{\vphantom{\frac{e^R}{R}}
    \left(\overline{\overline{\mathbf{I}}}_3-{\mathbf{e}}_\mathrm{R}  {\mathbf{e}}_\mathrm{R} \right)\right. +\left.\left(3 {\mathbf{e}}_\mathrm{R}   {\mathbf{e}}_\mathrm{R}  -\overline{\overline{\mathbf{I}}}_3\right)\left[\frac{1}{(k_0 R_{\alpha\beta})^{2}}-\frac{i}{k_0 R_{\alpha\beta}}\right]
    \right\},
    \label{Eq:g0}
\end{align}
where $k_0 = \omega_\mathrm{M}/c$. The separation vector is defined by $\mathbf{r}_\alpha - \mathbf{r}_\beta =\mathbf{R}_{\alpha\beta} = R_{\alpha\beta} {\mathbf{e}}_\mathrm{R} $ where ${\mathbf{e}}_\mathrm{R}$ is the unit vector and  $R_{\alpha\beta}=|\alpha-\beta|d$ is the distance along the x-axis.

In the presence of a silver surface (as shown in Figure~\ref{Fig_Schematic}), the dielectric function can be modeled by:
\begin{equation}
    \epsilon_\mathrm{r}(\mathbf{r},\omega) = 
    \begin{cases}
        1,     & \text{ if } z>0, \\
        \epsilon_\mathrm{Ag}(\omega),    & \text{ if } z\le 0,
    \end{cases} 
    \label{dielectrics}
\end{equation}
where the dielectric function of silver $\epsilon_\mathrm{Ag}(\omega)$ is obtained by fitting experimental data\cite{johnson_1972,Li1976} with a sophisticated model\cite{Melikyan2014} that ensures Kramers-Kronig relations\cite{scheel_1998}. Under this condition, the dyadic Green's function in eq~\ref{Eq:Green_Fun_full} can be decomposed into two components, $    \overline{\overline{\mathbf{G}}}(\mathbf{r}_\alpha,\mathbf{r}_\beta,\omega_\mathrm{M}) = \overline{\overline{\mathbf{G}}}_0(\mathbf{r}_\alpha,\mathbf{r}_\beta,\omega_\mathrm{M}) + \overline{\overline{\mathbf{G}}}_\mathrm{Sc}(\mathbf{r}_\alpha,\mathbf{r}_\beta,\omega_\mathrm{M})$, where $\overline{\overline{\mathbf{G}}}_\mathrm{Sc}(\mathbf{r}_\alpha,\mathbf{r}_\beta,\omega_\mathrm{M})$ denotes the scattering dyadic Green's function, arising from the presence of the silver surface. According to previous studies\cite{wu2018characteristic,Wang2020}, the dyadic Green's function of this planar layered dielectric structure can be computed using the Fresnel method\cite{chew1995waves,novotny2012principles}.
Note that, owing to the translational symmetry inherent in the arrangement of the chromophores and the surrounding dielectric environment, the dyadic Green's function depends only on the relative positions of $\mathbf{r}_\alpha$ and $\mathbf{r}_\beta$ and its distance from the silver surface, thus we can write $\overline{\overline{\mathbf{G}}}(\mathbf{r}_\alpha,\mathbf{r}_\beta,\omega_\mathrm{M}) = \overline{\overline{\mathbf{G}}}(\mathbf{r}_\alpha-\mathbf{r}_\beta,\omega_\mathrm{M}) = \overline{\overline{\mathbf{G}}}( R_{\alpha\beta},h,\omega_\mathrm{M})$.
In the end, the RDDI strength $V_{\alpha\beta}$ and the dissipation $\Gamma_{\alpha\beta}$ are determined by four parameters: the transition frequency $\omega_\mathrm{M}$, the amplitude of the transition dipole moment $\mu_\mathrm{M}$, the azimuthal angle $\varphi_\mathrm{M}$, and the intermolecular distance $R_{\alpha\beta}$. Furthermore, under these conditions, $\Delta_{\alpha}^{\text{Sc}}$ is identical for all molecules. Combined with the condition that all transition frequencies of chromophores are identical, we find that the relative small value of $\Delta_{\alpha}^{\text{Sc}}$ has a negligible effect on the quantum dynamics of chromophores' exciton state $\hat{\rho}_\mathrm{M}(t)$. Therefore, for simplicity, we neglect this term.

\begin{figure}[htbp] \includegraphics[width=0.99\textwidth]{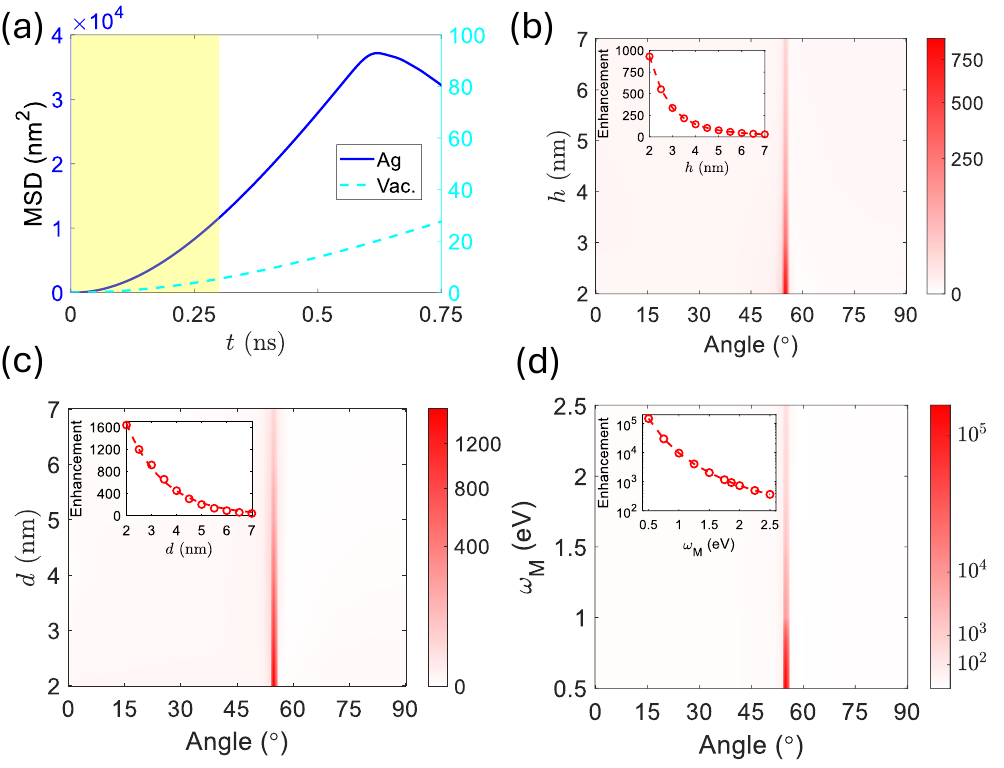}
\caption{ Time evolution of MSD and enhancement of diffusion coefficient of a molecular aggregate. (a) Time evolution of the MSD of the aggregate placed either in vacuum (cyan dashed line) or above the silver surface (blue solid line). The shaded yellow region highlights the time interval used to evaluate the diffusion coefficient of the exciton transport. A significant enhancement in exciton transport is observed when the aggregate is placed above the silver surface compared to vacuum.  
(b--d) Dependence of the enhancement factor of diffusion coefficients ($D/D_0$) on the azimuthal angle $ \varphi_\mathrm{M} $ in the $x$-axis and the indicated parameters in the $y$-axis:   
(b) chromophore-surface separation height $ h $,  
(c) nearest-neighbor intermolecular distance $ d $,  and
(d) chromophore transition frequency $\omega_\mathrm{M} $. 
Insets display the enhancement specifically at the magic angle $ \varphi_\mathrm{M} \approx 54.74^\circ $. We emphasize that the $y$-axis in panel (d) uses a logarithmic scale, showcasing the strong enhancement with respect to $ \omega_\mathrm{M} $. }
\label{Fig_Dynamics_Robust}
\end{figure}

To investigate non-equilibrium exciton transport, we simulate the dynamics of the mean square displacement (MSD) defined as follows\cite{sokolovskii_multi-scale_2023}:
\begin{align}
\label{Eq:MSD_Definition}
    \mathrm{MSD}(t) & \equiv \mathrm{Tr} \left[ \hat{\rho}_\mathrm{M}(t)  \sum_{\alpha=1}^{N_\mathrm{M}} (\hat{\mathbf{r}}_\alpha - \hat{\mathbf{r}}_1)^2   \right] \\ &=  d^2 \sum_{\alpha=1}^{N_\mathrm{M}} (\alpha-1)^2 P_\alpha(t),
\label{Eq:MSD_Numerical}
\end{align}
Here $P_{\alpha}(t)= \bra{\mathrm{X}_\alpha}  \hat{\rho}_\mathrm{M}(t) \ket{\mathrm{X}_\alpha}$ denotes the  population dynamics of the $\alpha$-th chromophore where $\ket{\mathrm{X}_\alpha} = \sigma^{(+)}_\alpha \ket{\mathrm{G}}$ is the single-excitation state and $\ket{\mathrm{G}}=\prod_\alpha\ket{\mathrm{g}_\alpha}$ is the collective ground state. The initial condition is assumed to be $\hat{\rho}_\mathrm{M}(0) = \ket{\mathrm{X}_1}\bra{\mathrm{X}_1}$. 
Based on eqs~\ref{Eq:GKSL_Eq} and \ref{Eq:MSD_Numerical}, the time evolution of MSD dynamics for the molecular aggregates is shown in Figure~\ref{Fig_Dynamics_Robust}a.

In contrast to suppressed exciton transport of a magic-angle-oriented molecular chain in a vacuum, the exciton transport dynamics near a metal surface are significantly enhanced due to non-vanishing intermolecular coupling mediated by radiative scattering. 
Figure~\ref{Fig_Dynamics_Robust}a shows that the MSD dynamics of the molecular aggregate exhibits a significantly faster growth when positioned at a height of $h=2$ nm above a silver surface (blue solid line) than its dynamics in a vacuum (cyan dashed line). Note that, due to the boundary reflection of the finite-size molecular chain, we observe an artificial turnover of the MSD dynamics around $ t=0.6\, \mathrm{ns}$ (see details in section S2 in SI). Therefore, we focus on the time evolution of the MSD for $t<0.3\ \mathrm{nm}$ (the shaded yellow region) and analyze its transport properties.

To further analyze the transport mechanism, we fit the MSD dynamics to the functional form\cite{xu_ultrafast_2023}:
\begin{align}
    \mathrm{MSD}(t) = 2D t^k
\label{Eq:MSD_fitting}
\end{align}
where $D$ represents the diffusion coefficient and $k$ is an exponent that dictates the exciton transport mechanism. By fitting the MSD data in Figure~\ref{Fig_Dynamics_Robust}a to eq~\ref{Eq:MSD_fitting}, we find that $k = 2$ for both in a vacuum and above a silver surface, suggesting that the exciton transport follows ballistic exciton transport. Moreover, the extracted diffusion coefficient values are $D \approx 2 \times 10^5~\mathrm{m^2/s^2}$ above the silver surface and $D_0 \approx 8 \times 10^1~\mathrm{m^2/s^2}$ for vacuum. Thus, an enhancement factor of three orders of magnitude ($D/D_0\sim 3 \times 10^3 $) is due to an increasing diffusion coefficient, rather than a different scaling with respect to time.

Next, we explore the robustness of this enhancement as observed in Figure~\ref{Fig_Dynamics_Robust}a against variations in several key parameters (while holding other parameters constant): the separation distance between the chromophores and the metallic surface $h$ in Figure~\ref{Fig_Dynamics_Robust}b, the intermolecular distance between nearest-neighbor chromophores $d$ in Figure~\ref{Fig_Dynamics_Robust}c, and the transition frequency of chromophores $\omega_\mathrm{M}$ in Figure~\ref{Fig_Dynamics_Robust}d. Overall, we observe that a robust enhancement peak occurs consistently near the magic angle ($\varphi_\mathrm{M} \approx 54.74^\circ$) while the enhancement factor decreases as $h$, $d$, or $\omega_\mathrm{M}$ increases.
Focusing on the magic angle, the specific decay of the enhancement factor implies various parameter dependence of exciton transport mediated by radiative scattering.
First, when varying $h$ (with $d=3~\mathrm{nm}$ and $\omega_\mathrm{M}=1.864~\mathrm{eV}$ constant), the insert in Figure~\ref{Fig_Dynamics_Robust}b shows a decay of $D/D_0\propto h^{-2.5}$, highlighting the near-surface nature of the exciton transport enhancement. 
Second, for variations in $d$ (holding $h=2~\mathrm{nm}$ and $\omega_\mathrm{M}=1.864~\mathrm{eV}$ constant), the insert in Figure~\ref{Fig_Dynamics_Robust}c shows a decay of $D/D_0\propto d^{-2.0}$. This observed scaling is close to the $d^{-3}$ scaling of the near-field dipole-dipole interactions, suggesting a strong link to intermolecular coupling.
Finally, when $\omega_\mathrm{M}$ is varied (holding $h=2~\mathrm{nm}$ and $d=3~\mathrm{nm}$ constant), the insert of Figure~\ref{Fig_Dynamics_Robust}d shows a decay of $D/D_0\propto \omega_\mathrm{M}^{-3.9}$, indicating that the enhancement is more pronounced for molecules with lower transition frequencies of chromophores when $\omega_\mathrm{M}$ is far off-resonant with the plasmonic frequency of the silver surface (approximately $3.6~\mathrm{eV}/\hbar$).

To elucidate the underlying mechanism of these observed scaling using the MQED framework, we analyze the fundamental quantities obtained in the MQED calculations, including the nearest-neighbor RDDI coupling strength ($V_{\alpha\beta}$) and dissipation rate ($\Gamma_{\alpha\beta}$) where $\beta = \alpha+1$.
Specifically, we focus on the ratio $V_{\alpha\beta}/V_{0,\alpha\beta}$ and $\Gamma_{\alpha\beta}/\Gamma_{0,\alpha\beta}$ where $V_{0,\alpha\beta}$ and $\Gamma_{0,\alpha\beta}$ are the corresponding quantities in a vacuum (i.e., $\overline{\overline{\mathbf{G}}}(\mathbf{r}_\alpha,\mathbf{r}_\beta,\omega_\mathrm{M})$ in eqs~\ref{Eq:RDDI_Definition} and \ref{Eq:Gamma_Definition} is substituted with $\overline{\overline{\mathbf{G}}}_0(\mathbf{r}_\alpha,\mathbf{r}_\beta,\omega_\mathrm{M})$, as defined in eq~\ref{Eq:g0}). Again, these ratios are explored with respect to key parameters: chromophore-surface separation $h$, nearest-neighbor intermolecular distance $d$, and transition frequency of chromophores $\omega_\mathrm{M}$. 

\begin{figure}[htbp] \includegraphics[width=1\textwidth]{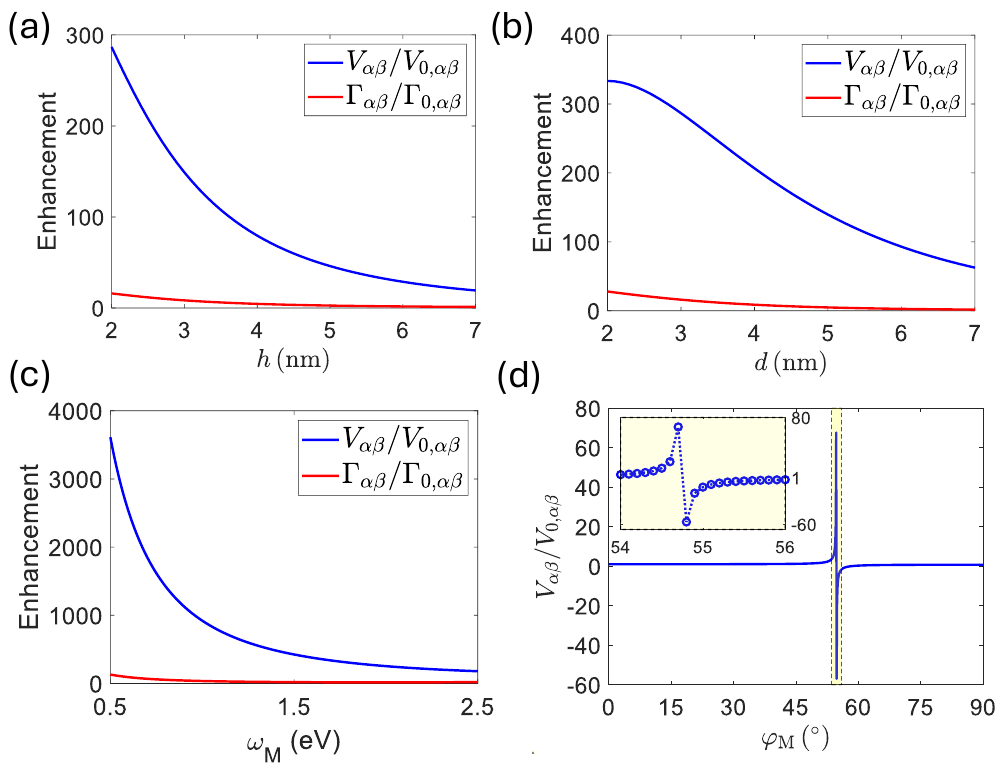}
\caption{ Enhancement ratios of the resonant dipole-dipole interaction (RDDI), represented by $V_{\alpha\beta}/V_{0,\alpha\beta}$ (blue solid lines), and the dissipation, given by $\Gamma_{\alpha\beta}/\Gamma_{0,\alpha\beta}$ (red solid lines), are plotted as functions of:
(a) chromophore-surface separation $h$,
(b) nearest-neighbor intermolecular distance $d$,
(c) transition frequency of the chromophores $\omega_\mathrm{M}$, and
(d) azimuthal angle $\varphi_\mathrm{M}$.}
\label{Fig_Exploration}
\end{figure}

In Figure~\ref{Fig_Exploration}, we show that the silver surface substantially amplifies $V_{\alpha\beta}$ by approximately $10^2-10^3$ fold compared to its vacuum counterpart $V_{0,\alpha\beta}$, directly demonstrating it facilitates exciton migration. Notably, we find that the decay trend of $V_{\alpha\beta}/V_{0,\alpha\beta}$ closely agrees with the corresponding scaling of the diffusion coefficient enhancement $D/D_0$ as observed in Figure~\ref{Fig_Dynamics_Robust}. 
First, Figure~\ref{Fig_Exploration}a shows $V_{\alpha\beta}/V_{0,\alpha\beta}\propto h^{-2}$, indicating a predominantly near-surface effect similar to $D/D_0\propto h^{-2.5}$. 
Second, in the $d>3.5~\mathrm{nm}$ regime, Figure~\ref{Fig_Exploration}b shows $V_{\alpha\beta}/V_{0,\alpha\beta}\propto d^{-2}$ consistent with $D/D_0\propto d^{-2.0}$ shown in Figure~\ref{Fig_Dynamics_Robust}c inset, reflecting the enhanced scaling with respect to the intermolecular coupling. Below $d = 3~\mathrm{nm}$, the influence of $h$ on the ratio $V_{\alpha\beta}/V_{0,\alpha\beta}$ becomes pronounced, causing it to deviate from the $d^{-2}$ dependence.
Finally, we notice that, with increasing $\omega_\mathrm{M}$, Figure~\ref{Fig_Exploration}c shows $V_{\alpha\beta}/V_{0,\alpha\beta}\propto \omega_\mathrm{M}^{-2}$, different from the scaling behavior of $D/D_0\propto\omega_\mathrm{M}^{-3.9}$.
This discrepancy implies that the functional dependence of the diffusion coefficient $D$ with respect to the RDDI coupling strength $V$ can be influenced by the presence of the radiative scattering via the metallic surface.
These quantitative analysis suggests that the enhanced RDDI coupling is responsible for the diffusion coefficient enhancement. 

In contrast, the dissipation rate enhancement $\Gamma_{\alpha\beta}/\Gamma_{0,\alpha\beta}$ is much smaller than the RDDI coupling enhancement. As shown in Figures~\ref{Fig_Exploration}a-\ref{Fig_Exploration}c, the maximal enhancement is $\Gamma_{\alpha\beta}/\Gamma_{0,\alpha\beta}\sim20$ and invariably decays as $h$, $d$, or $\omega_\mathrm{M}$ increases. This observation confirms that the metallic surface primarily enhances exciton transport through the enhanced RDDI, and the impact on the dissipation rates is comparatively negligible. 
Note that these analyses are conducted under the condition that $\omega_\mathrm{M}$ is less than silver's surface plasmon polariton frequency (approximately $3.6~\mathrm{eV}/\hbar$).

Now we turn our attention to the azimuthal angle dependence of the RDDI coupling.
As shown in Figure~\ref{Fig_Exploration}d, the RDDI enhancement ratio generally remains close to unity ($V_{\alpha\beta}/V_{0,\alpha\beta}\approx 1$) across most of the angular range. However, it displays sharp singularities within a narrow critical region around $\varphi_\mathrm{M} \approx 54^\circ\text{--}56^\circ$ (highlighted by a yellow shaded box), showing a rapid transition from large positive values to large negative values.
Such a phase flip of the RDDI enhancement corresponds to the transition from J-aggregates to H-aggregates.

\begin{figure}[htbp] \includegraphics[width=0.99\textwidth]{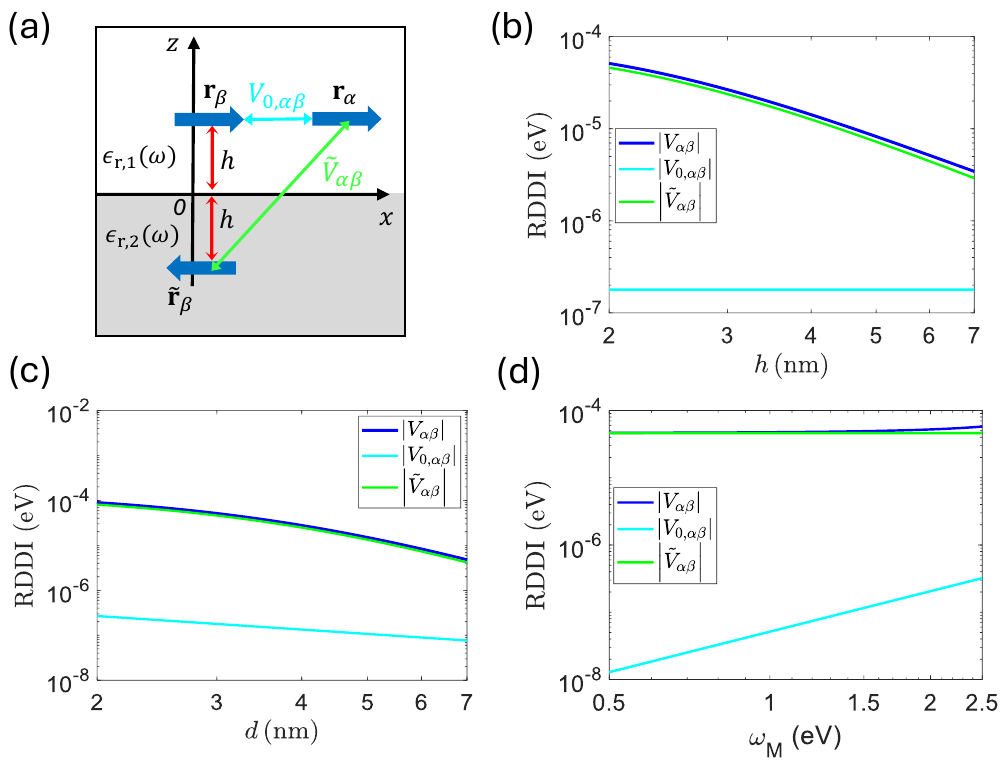}
\caption{ Schematic diagram of the image-dipole method and scaling dependence of resonant dipole-dipole interactions (RDDI).
(a) Schematic illustration of the image-dipole method. The numerically exact RDDI between two chromophores ($\alpha$ and $\beta$) positioned above a silver surface ($V_{\alpha\beta}$), calculated via MQED, is approximated as the sum of their direct vacuum interaction ($V_{0,\alpha\beta}$) and the image-mediated interaction ($\tilde{V}_{\alpha\beta}$) between the acceptor chromophore at $\mathbf{r}_\alpha$ and the donor's image dipole at $\tilde{\mathbf{r}}_\beta$.
(b) Log-log plot of the RDDI magnitude as a function of chromophore-surface separation $h$. Curves represent the numerically exact RDDI ($|V_{\alpha\beta}|$, blue), the direct vacuum interaction ($|V_{0,\alpha\beta}|$, cyan), and the image-mediated interaction ($|\tilde{V}_{\alpha\beta}|$, green).
(c) Log-log plot of the RDDI magnitude as a function of nearest-neighbor intermolecular distance $d$.
(d) Log-log plot of the RDDI magnitude as a function of the chromophore transition frequency $\omega_\mathrm{M}$. The excellent agreement between $V_{\alpha\beta}$ and $\tilde{V}_{\alpha\beta}$ observed in panels (b-d) validates the image-dipole method across the investigated ranges of $h$, $d$, and $\omega_\mathrm{M}$. This agreement further demonstrates the predominance of the image-mediated term over the direct vacuum component in determining $V_{\alpha\beta}$ for molecular aggregates oriented at the magic angle. }
\label{Fig_Image}
\end{figure}

Finally, to further elucidate the approximate inverse-square scaling of the RDDI enhancement observed in Figures~\ref{Fig_Exploration}a-\ref{Fig_Exploration}c with respect to $h$, $d$, and $\omega_\mathrm{M}$, we employ the image-dipole method \cite{jackson2021classical} to gain a qualitative insight into the RDDI coupling. 
As illustrated in Figure~\ref{Fig_Image}a, the RDDI coupling strength $V_\mathrm{\alpha\beta}$ between the $\alpha$-th and $\beta$-th chromophores calculated based on MQED can be approximated by the sum of the direct vacuum interaction, $V_\mathrm{0,\alpha\beta}$, and the image-mediated interaction $\tilde{V}_{\alpha\beta}$ between the $\alpha$-th chromophore and the image of the $\beta$-th chromophore: $V_\mathrm{\alpha\beta} \approx V_\mathrm{0,\alpha\beta} + \tilde{V}_{\alpha\beta}$, where $\tilde{V}_{\alpha\beta}$ is given by
\begin{align}\label{RDDI_Image}
    \tilde{V}_{\alpha\beta} = \frac{- \omega_\mathrm{M}^2}{ \epsilon_0 c^2}  \boldsymbol{\mu}_{\alpha} \cdot \mathrm{Re} \overline{\overline{\mathbf{G}}}_0(\mathbf{r}_\alpha,\tilde{\mathbf{r}}_\beta, \omega_\mathrm{M}) \cdot \tilde{\boldsymbol{\mu}}_\beta.
\end{align}
Here, given that the $\beta$-th chromophore is located at $\mathbf{r}_\beta = (0,0,h)$ with dipole moment $\boldsymbol{\mu}_\beta = \mu_\mathrm{M}(\cos\varphi_\mathrm{M},\sin\varphi_\mathrm{M},0)$ , its image dipole is at $\tilde{\mathbf{r}}_\beta = (0,0,-h)$, with the corresponding dipole moment $\tilde{\boldsymbol{\mu}}_\beta = \mu_\mathrm{M}(-\cos\varphi_\mathrm{M},-\sin\varphi_\mathrm{M},0) $.
To validate the image-dipole method, we confirm the corresponding RDDI enhancements $V_{\text{Imag},\alpha\beta}/V_{\text{0},\alpha\beta}$ with respect to $h$, $d$, and $\omega_\mathrm{M}$ agree with the MQED results in Figure~\ref{Fig_Exploration} (see details in section S3 in SI).

As depicted in Figures~\ref{Fig_Image}b-\ref{Fig_Image}d, the coupling strength contributed by the image dipole $\tilde{V}_{\alpha\beta}$ is considerably larger than the direct vacuum interaction $V_\mathrm{0,\alpha\beta}$, indicating that the surface-induced interaction predominantly contributes to the total interaction, i.e., ${V}_{\alpha\beta}\approx \tilde{V}_{\alpha\beta} $ for $d, \, h < 7\, \mathrm{nm}$. Furthermore, the close agreement observed in these figures between the blue lines (result based on MQED) and the green lines (representing the image-dipole model) suggests that the silver surface behaves effectively as a perfect reflecting mirror for evaluating $V_\mathrm{\alpha\beta}$ in the near-infrared and visible light regions ($0.5\,\mathrm{eV}<\hbar\omega_\mathrm{M}<2.5\, \mathrm{eV}$).

Let's further analyze the scaling of $V_\mathrm{0,\alpha\beta}$ and $\tilde{V}_{\alpha\beta}$ based on the image dipole model. 
First, we find that the direct vacuum interaction $V_\mathrm{0,\alpha\beta}$ exhibits a $d^{-1}$ dependence (Figure~\ref{Fig_Image}c, cyan line), rather than the conventional $d^{-3}$ scaling. 
This is because the orientation factor $\kappa = (\boldsymbol{\mu}_\alpha \cdot \boldsymbol{\mu}_\beta - 3(\boldsymbol{\mu}_\alpha\cdot \mathbf{R}_{\alpha\beta})(\boldsymbol{\mu}_\beta\cdot \mathbf{R}_{\alpha\beta}) )/(\abs{\boldsymbol{\mu}_\alpha}\abs{\boldsymbol{\mu}_\beta}\abs{\mathbf{R}_{\alpha\beta}}^2)=0 $ at the magic-angle configuration ($\varphi_\mathrm{M} = \arccos(1/\sqrt{3})$) where $\mathbf{R}_{\alpha\beta}=\mathbf{r}_\alpha-\mathbf{r}_\beta = \mathbf{e}_\mathrm{R} \abs{\alpha-\beta}d $. Thus, the $d^{-3}$ dependence in the RDDI coupling strength is eliminated (see section S4 in the SI).
Second, the interaction caused by the image dipole $\tilde{V}_{\alpha\beta}$ scales as $\tilde{R}_{\alpha\beta}^{-3}$, where $\tilde{R}_{\alpha\beta}= \sqrt{ d^2+4h^2 }$ is the distance between the $\alpha$-th chromophore and the image dipole (see section S5 in the SI). This implies that $V_{\alpha\beta}\approx \tilde{V}_{\alpha\beta}\sim d^{-3}$ and $V_{\alpha\beta}\approx\tilde{V}_{\alpha\beta} \sim h^{-3}$ when $d$ is comparable to $h$, as shown in Figures~\ref{Fig_Image}b and \ref{Fig_Image}c. 
Therefore, the observed $d^{-2}$ or $h^{-2}$ scaling for the enhancement factor $V_{\alpha\beta}/V_{0,\alpha\beta}$ can be explained by $\tilde{V}_{\alpha\beta}/V_{0,\alpha\beta}\sim d^{-3}/d^{-1}$.

Finally, regarding the $\omega_\mathrm{M}$ dependence, we find that the image-dipole method demonstrates that $\tilde{V}_{\alpha\beta} \sim \omega_\mathrm{M}^0$ (Figure~\ref{Fig_Image}d, green line) in the quasi-static limit ($\omega_\mathrm{M} \tilde{R}_{\alpha\beta}/c \ll 1$), whereas $V_{0,\alpha\beta} \sim \omega_\mathrm{M}^2$ (Figure~\ref{Fig_Image}d, cyan line). This $\omega_\mathrm{M}^2$ dependence arises because, at the magic-angle configuration, the dominating $d^{-1}$ term of $\overline{\overline{\mathbf{G}}}_0(\mathbf{r}_\alpha,\mathbf{r}_\beta, \omega_\mathrm{M})$ in eq~\ref{Eq:g0} is independent of $\omega_\mathrm{M}$ and eq~\ref{Eq:RDDI_Definition} scales as $\omega_\mathrm{M}^2$, leading to an overall scaling of $V_{0,\alpha\beta} \sim \omega_\mathrm{M}^2$ (see also eq~S28 in section S4 in SI). 
On the other hand, the $\tilde{V}_{\alpha\beta}$ is dominated by the near-field coupling term (the $d^{-3}$ term) in eq~\ref{Eq:g0} (see also eq S38 in section S5 in SI), so that $\overline{\overline{\mathbf{G}}}_0(\mathbf{r}_\alpha,\tilde{\mathbf{r}}_\beta, \omega_\mathrm{M})\sim\omega_\mathrm{M}^{-2}$ (note that $k_0 = \omega_\mathrm{M}/c$) cancel the $\omega_\mathrm{M}^{2}$ dependence of $\tilde{V}_{\alpha\beta}$ in eq~\ref{RDDI_Image}. Consequently, the $\omega_\mathrm{M}^{-2}$ scaling for the enhancement factor $V_{\alpha\beta}/V_{0,\alpha\beta}$ as observed in Figure~\ref{Fig_Exploration}c can be understood in terms of the image-dipole model. 
We emphasize that, at the magic-angle orientation, $V_{0,\alpha\beta}$ is solely due to long-range dipole-dipole interactions (i.e., the $\omega_\mathrm{M}^2d^{-1}$ scaling term) while $\tilde{V}_{\alpha\beta}$ is contributed by the near-field coupling (i.e., the $\omega_\mathrm{M}^0d^{-3}$ scaling term). Interestingly, similar to the $\omega_\mathrm{M}$ scaling of the RDDI coupling, the diffusion coefficient follows $D\sim\omega_\mathrm{M}^{0.2}$ and $D_0\sim\omega_\mathrm{M}^4$ (see details in section S6 in SI). This implying that, in proximity to a metallic surface, both the diffusion coefficient ($D$) and the RDDI coupling ($V_{\alpha\beta}$) exhibit robustness against variations in molecular transition frequency. In other words, the enhanced exciton transport mediated by the radiative scattering of the metallic surface remains a constant for various transition frequencies. This intriguing phenomenon offers a possible route for experimental examination.


In conclusion, we investigate exciton transport in molecular aggregates with a magic-angle orientation---a configuration that has been largely overlooked despite its unique capacity to bridge transport theory and light-matter interaction studies. While chromophore arrays in free space exhibit negligible RDDI at magic-angle orientations, we demonstrate that coupling these systems to a metallic surface results in significant enhancement of exciton transport, thereby directly addressing the primary research question. Moreover, concerning the second inquiry, the study reveals that this enhancement is robust regardless of variations in molecular transition frequencies $\omega_\mathrm{M}$, nearest-neighbor intermolecular distances $d$, and chromophore-surface separations $h$, underscoring the potential of magic-angle configurations to enable tunable exciton dynamics in hybrid photonic or plasmonic environments. Finally, regarding the third question, our numerical simulations show that the enhancement of the diffusion coefficient ($D_{\alpha\beta}/D_{0,\alpha\beta}$) exhibits approximate scaling dependencies of $\sim h^{-2.5}$, $\sim d^{-2.0}$, and $\sim \omega_\mathrm{M}^{-3.9}$. These trends should be closely related to the scaling behavior of the RDDI enhancement ($V_{\alpha\beta}/V_{0,\alpha\beta}$), which scales approximately as $\sim h^{-2}$, $\sim d^{-2}$, and $\sim \omega_\mathrm{M}^{-2}$. The image-dipole method offers a qualitative physical explanation for this scaling behavior of the RDDI enhancement.

For the purpose of experimentally validating the predicted enhancement of exciton transport, we propose exploring systems similar to CF$_3$DPT solids\cite{ghosh_2024}, contingent upon their intermolecular distances exceeding 1 nm. Distances of lesser magnitude (e.g., $d < 1~\mathrm{nm}$) would engender Dexter-type exchange coupling, thereby rendering the predictions herein presented nugatory. If finding suitable molecular aggregates is difficult, alternative systems may be considered, including, but not limited to, metal-organic frameworks (MOFs), covalent organic frameworks (COFs), or semiconductor quantum-emitter arrays\cite{palaciosberraquero_2017}. As a complementary validation strategy, in case direct observation of exciton transport remains experimentally challenging, we propose that RDDI be prioritized for initial experimental characterization. Preceding investigations have demonstrated that the measurement of mutual impedance between two electric dipoles at microwave frequencies enables direct access to the complete dyadic Green's function\cite{rustomji_2021,lezhennikova_2023}. Based on eqs~\ref{Eq:RDDI_Definition} and \ref{Eq:Gamma_Definition}, the corresponding $V_{\alpha\beta}$ and $\Gamma_{0,\alpha\beta}$ can be computed directly. These experimental strategies, employed in conjunction with the theoretical framework presented herein, may offer a comprehensive approach to elucidating the interplay between the metallic surface and exciton transport within magic-angle-oriented molecular aggregates. We hope that this study could motivate further experimental and theoretical investigations into the fundamental tenets of transport theory, as well as the dynamics of light-matter interactions in complex dielectric environments.

\begin{suppinfo}

Derivation of dynamic equation eq~1; Population dynamics of magic-angle-oriented molecular aggregate; Comparison of RDDI enhancements: image-dipole method vs. MQED results; Expression of $V_{0,\alpha\beta}$ and $\Gamma_{0,\alpha\beta}$ for magic-angle-oriented molecular aggregates in free space; Expression of $V_{\mathrm{Imag},\alpha\beta}$ and $\Gamma_{\mathrm{Imag},\alpha\beta}$ based on image-dipole method; Scaling behavior of diffusion coefficient in vacuum and above silver surface.  

\end{suppinfo}

\begin{acknowledgement}
Wang and Chen thank the support provided by the University of Notre Dame and the Asia Research Collaboration grant of Notre Dame Global. Wang and Hsu also thank Academia Sinica (AS-CDA-111-M02), National Science and Technology Council (111-2113-M-001-027-MY4), and Physics Division, National Center for Theoretical Sciences (112-2124-M-002-003) for the financial support.

\end{acknowledgement}

\bibliography{References}

\end{document}